\documentclass[fleqn,usenatbib]{mnras}

\usepackage{newtxtext,newtxmath}
\usepackage{bm}
\usepackage{enumitem}
\usepackage{makecell}
\usepackage{lineno}
\hypersetup{
	colorlinks = true,
	linkcolor = blue,
	anchorcolor = blue,
	citecolor = blue,
	urlcolor = blue
}
\usepackage[caption=false]{subfig}
\usepackage{soul, color, xcolor}
\usepackage{multirow}
\usepackage{booktabs}
\usepackage[T1]{fontenc}

\DeclareRobustCommand{\VAN}[3]{#2}
\let\VANthebibliography\thebibliography
\def\thebibliography{\DeclareRobustCommand{\VAN}[3]{##3}\VANthebibliography}

\usepackage{graphicx}	% Including figure files
\usepackage{amsmath}	% Advanced maths commands

\title[Poynting-Flux-Dominated Outflow of GRB 230307A]{The Jet Composition of GRB 230307A: Poynting-Flux-Dominated Outflow?}

\author[Du et al.]{
Zhao-Wei Du,$^{1}$
HouJun L\"{u},$^{1}$\thanks{E-mail: lhj@gxu.edu.cn}
Xiaoxuan Liu$^{1}$
and EnWei Liang$^{1}$
\\
$^{1}$Guangxi Key Laboratory for Relativistic Astrophysics, Department of Physics, Guangxi University, Nanning 530004, China;
}

\date{Accepted XXX. Received YYY; in original form ZZZ}

\pubyear{2023}

\begin{document}
\label{firstpage}
\pagerange{\pageref{firstpage}--\pageref{lastpage}}
\maketitle

\begin{abstract}
The jet composition of GRB plays an important role in understanding the energy dissipation and radiation mechanisms in GRB physics, but it is poorly constrained from the observational data. Recently, an interesting long-duration GRB 230307A with redshift $z=$0.065 has attracted great attention. The lack of detected thermal emission and mini-structure of prompt emission lightcurve of this burst suggest that the outflow is Poynting-flux-dominated and point towards the ICMART model. In this paper, we invoke two independent methods to investigate the jet composition of GRB 230307A. The high magnetization parameter ($\sigma>7$ or ever large) for$R_0=10^{10}$ cm that is used to suppress thermal component, strongly suggests that a significant fraction of the outflow energy is likely in a Poynting flux entrained with the baryonic matter. Moreover, it is found that the radiation efficiency of this burst for typical values $\epsilon_e=0.1$ and $\epsilon_B=0.01$ can reach as high as $~50\%$ which disfavors the internal shock model, but is consistent with ICMART model. Finally, a possible unified picture to produce GRB 230307A originated from a compact star merger is also discussed.  
\end{abstract}


%%%%%%%%%%%%%%%%%%%%%%%%%%%%%%%%%%%%%%%%%%%%%%%%%%

%%%%%%%%%%%%%%%%% BODY OF PAPER %%%%%%%%%%%%%%%%%%

\section{Introduction} \label{sec:intro}
The observed prompt emission of Gamma-Ray burst (GRB) is believed to be from an ultra-relativistic jet, which is launched from the central engine during a catastrophic event, e.g., the core collapse of a massive star or merger of two compact stars \citep{2015PhR...561....1K,2018pgrb.book.....Z}. Although the prompt emission of GRB was discovered much earlier than the afterglow, several open questions remain poorly to be understood in GRB physics, e.g., what is the composition of a GRB jet? How energy is dissipated to give rise to prompt emission? \citep{2011CRPhy..12..206Z}.

Traditionally, two scenarios of jet composition have been proposed to interpret the observations of GRB. One is matter-dominated fireball which is composed of hot photons, electron/positron pairs, and a small amount of baryons \citep{1986ApJ...308L..43P,1986ApJ...308L..47G,1990ApJ...365L..55S}. An initial fireball is accelerated to a relativistic speed under its own thermal pressure, and a fraction of the thermal energy is converted to the kinetic energy of the outflow \citep{1993ApJ...415..181M,1993MNRAS.263..861P}. A quasi-thermal component from the fireball photosphere should be powered when the fireball becomes transparent at the photosphere radius (photosphere model; \citealt{1986ApJ...308L..43P,1986ApJ...308L..47G}). Moreover, a fraction of the kinetic energy of the outflow is further dissipated into heat and radiation in internal shocks to produce a non-thermal component at  a larger radius (internal shock model; \citealt{1994ApJ...430L..93R,1997ApJ...490...92K,1998MNRAS.296..275D,2006ApJ...642..995P,2014IJMPD..2330002Z}). An alternative scenario invokes a non-thermal component from the Poynting-flux-dominated outflow where most of the energy is stored in the magnetic field. The magnetic energy can be dissipated through magnetic reconnection or current instability to power the observed prompt emission of GRB \citep{1994MNRAS.270..480T,2003astro.ph.12347L,2009ApJ...700L..65Z}, or the internal-collision-induced magnetic reconnection and turbulence (ICMART) model \citep{2011ApJ...726...90Z}. 

From the theoretical point of view, the photosphere and internal shock models have a magnetization parameter $\sigma$ much less than unity at the GRB emission site, while the ICMART model has a moderately large $\sigma>1$ at the emission site, with the GRB emission powered by directly dissipating the magnetic energy to radiation \citep{2011ApJ...726...90Z,2015ApJ...801..103G,2023ApJ...943..146C}. Also, the internal shock model requires a high radiation efficiency in the prompt emission \citep{1999ApJ...523L.113K,1999ApJ...522L.105P}.

Observationally, (1) the quasi-thermal component predicted by the fireball model was observed in GRB 090902B \citep{2009ApJ...706L.138A,2010ApJ...709L.172R,2012MNRAS.420..468P}, and it implies the matter-dominated outflow of GRB jet; (2) A good fraction of GRBs are consistent with not having a thermal component, e.g., GRB 080916C \citep{2009Sci...323.1688A,2009ApJ...700L..65Z} and GRB 130606B \citep{2016ApJ...816...72Z,2017ApJ...846..137O,2019A&A...625A..60R,2020NatAs...4..174B}, and it suggests the Poynting-flux-dominated outflow of GRB jet; (3) A dominant non-thermal component and a sub-dominant thermal component have been discovered, e.g., GRB 100724B \citep{2011ApJ...727L..33G}, GRB 110721A \citep{2012ApJ...757L..31A}, GBR 160625B \citep{2017ApJ...849...71L,2017Natur.547..425T,2018NatAs...2...69Z}, GRB 081221 \citep{2018ApJ...866...13H}, and 211211A \citep{2023ApJ...943..146C}, which make us believe the “hybrid” jet of GRB. In any case, the rich data observed by Fermi Gamma-Ray Burst Monitor (GBM) suggest that the GRB jet composition is likely diverse. However, it is still difficult to diagnose the jet composition of most GRBs from observational data \citep{2018pgrb.book.....Z}. 

Recently, an interesting long-duration GRB 230307A with redshift $z=$0.065 that triggered the Fermi/GBM \citep{2023GCN.33407....1D}, Gravitational wave high-energy Electromagnetic Counterpart All-sky Monitor (GECAM; \cite{2023GCN.33406....1X}), as well as Konus-Wind \citep{2023GCN.33427....1S}, is very excited for attention and follow up observations by other telescopes \citep{2023ApJ...954L..29D,2023arXiv230702098L,2023arXiv230800638Y,2023arXiv230705689S,2023arXiv231007205Y}. The properties of the light curve are quite similar to that of GRB 060614 \citep{2006Natur.444.1044G}, GRB 211211A \citep{2022Natur.612..228T,2022Natur.612..223R,2022Natur.612..232Y,2023ApJ...943..146C}, and GRB 211227A \citep{2022ApJ...931L..23L,2023A&A...678A.142F}. No associated supernova signature, but a possible association with kilonova, together with heavy element nucleosynthesis, suggest that GRB 230307A originated from a binary compact star merger event \citep{2023ApJ...954L..29D, 2023arXiv230702098L, 2023arXiv230800638Y, 2023arXiv230705689S,2023arXiv231007205Y}. More interestingly, \cite{2023arXiv231007205Y} claimed that the light curve of prompt emission observed by GECAM is composed of many rapidly variable short pulses. They suggested that the jet composition is Poynting-flux-dominated and is consistent with the ICMART model.

In this paper, we invoke an independent method to investigate the jet composition of GRB 230307A. The lack of detection of a suppressed thermal component, and high magnetization parameter $\sigma$, together with high radiation efficiency, also strongly suggest a Poynting-flux-dominated jet of GRB 230307A which is consistent with the results of \cite{2023arXiv231007205Y}. The constraints from the suppressed thermal component are presented in \S2. The detail for calculating radiation efficiency is shown in \S3.  Conclusions are drawn in \S 4 with some additional discussions. Throughout the paper, we use the notation $Q=10^{n}Q_{n}$ in CGS units and adopt a concordance cosmology with parameters $H_{\rm 0}=71~\rm km~s^{-1}~Mpc^{-1}$, $\Omega_{\rm M}=0.30$, and $\Omega_{\Lambda}=0.70$.

\section{Constraints from lack of detected thermal emission} \label{sec:BB}
Based on the fireball model, the emission site at which the fireball becomes transparent, is called the photosphere radius when the electron scattering optical depth ($\tau_{\gamma e}^{\prime}=n^{\prime}\sigma_T \Delta^{\prime}$) is close to 1. Here, $\sigma_T$ is the Thomson cross section, $n^{\prime}$ and $\Delta^{\prime}$ are electron number density and width of the ejecta shell in the rest frame comoving with the ejecta, respectively \citep{2000ApJ...530..292M, 2005ApJ...628..847R, 2007ApJ...666.1012T, 2007MNRAS.382L..72G, 2009ApJ...700L..47L}. By assuming a pure baryonic flux, we derive a thermal component spectrum that can be emitted from the photosphere with a total wind luminosity \textit{of} $L_{\rm w}$ \citep{2018pgrb.book.....Z}. Following the method in \cite{2009ApJ...700L..65Z}, the photosphere radius can be written as \citep{2000ApJ...530..292M, 2008ApJ...682..463P, 2015ApJ...801..103G, 2018pgrb.book.....Z}
\begin{equation}
\begin{split}
\label{111}
R_{\rm ph}=\left\{
\begin{aligned}
&\Big(\frac{L_{\rm w} \sigma_{\rm T} R_0^2}{8\pi m_{\rm p} c^3\eta}\Big)^{1/3},&R_{\rm ph}<R_{\rm c}\\
&\frac{L_{\rm w}\sigma_{\rm T}}{8\pi m_{\rm p}c^3\Gamma^2\eta}, &R_{\rm ph} > R_{\rm c}
\end{aligned}
\right.
\end{split}
\end{equation}
where $\eta=L_{\rm w}/\dot Mc^2$ is dimensionless entropy of baryonic flow, $R_{\rm c}\sim R_0\times \rm min(\eta,\eta_*)$ is the radius where ejecta enter the 'coasting' phase, $R_0=c\delta t^{\rm ob}$ is the radius at which the ejecta is emitted from central engine, $\eta_*=(L_{\rm w}\sigma_T/8\pi m_{\rm p}c^3R_0)^{1/4}$ is critical dimensionless entropy, $t^{\rm ob}$ is the variability time scale of the central engine, $m_p$ and $c$ are the fundamental constants proton mass and speed of light, respectively. The coasting Lorentz factor is $\Gamma=\eta$ and $\Gamma=\eta_*$ for $R_{\rm ph}>R_{\rm c}$ and $R_{\rm ph}\leq R_{\rm c}$, respectively. 

Observed gamma-ray luminosity $L_{\gamma}$ should be below initial wind luminosity $L_{\rm w}$ of fireball, i.e., $L_{\rm w}\geq L_{\gamma}$. This kind of outflow contains large residual energy which may be released at the photosphere radius. So that the luminosity of the thermal component can be written as \citep{2000ApJ...530..292M}:
\begin{equation}
\begin{split}
\label{9}
L_{\rm th}=\left\{
\begin{aligned}
&L_{\rm w}, & \eta > \eta_*,~R_{\rm ph} < R_{\rm c}\\
&L_{\rm w}(\eta/\eta_*)^{8/3}, & \eta < \eta_*,~R_{\rm ph} > R_{\rm c}
\end{aligned}
\right.
\end{split}
\end{equation}
One can calculate the temperature of the blackbody component which is produced from the photosphere \citep{2000ApJ...530..292M, 2008ApJ...682..463P},
\begin{equation}
\begin{split}
\label{9}
T^{\rm ob}_{\rm ph}=\left\{
\begin{aligned}
&(\frac{L_{\rm w}}{4\pi R_0^2 a})^{1/4}(1 + z)^{-1}, & R_{\rm ph}<R_{\rm c}\\
&(\frac{L_{\rm w}}{4\pi R_0^2 a})^{1/4}(\frac{R_{\rm ph}}{R_{\rm c}})^{-2/3}(1 + z)^{-1}, & R_{\rm ph}>R_{\rm c}
\end{aligned}
\right.
\end{split}
\end{equation}
where $a$ is Stefan-Boltzman's constant. 

Observationally, from soft X-rays to gamma-rays, \cite{2023arXiv230705689S} performed a detailed joint-spectral analysis of the data of GRB 230307A that is observed by both GECAM and LEIA. They found that the joint spectral energy distribution (SED) can be described by power-law model, cutoff power-law model, or BAND-Cut model without thermal emission \citep{2023arXiv230705689S}. Furthermore, the theoretically expected thermal peak energy is below $\sim 50$ keV, but the photons can be heavily absorbed by column density of hydrogen below 1 keV. In order to make the coverage of energy range between 1 keV and 50 KeV, we select the spectral fitting results of the time interval for five epochs, e.g., [13-18]s, [18-25]s, [25-35]s, [35-50]s, and [50-75]s, and mark those five epochs as (a), (b), (c), (d), and (e), respectively. One interesting question is how strong of the outflow is magnetization, and it makes the lack of detection of thermal component which is suppressed.

\begin{figure*}
%    \centering
    \includegraphics[width=3.5in]{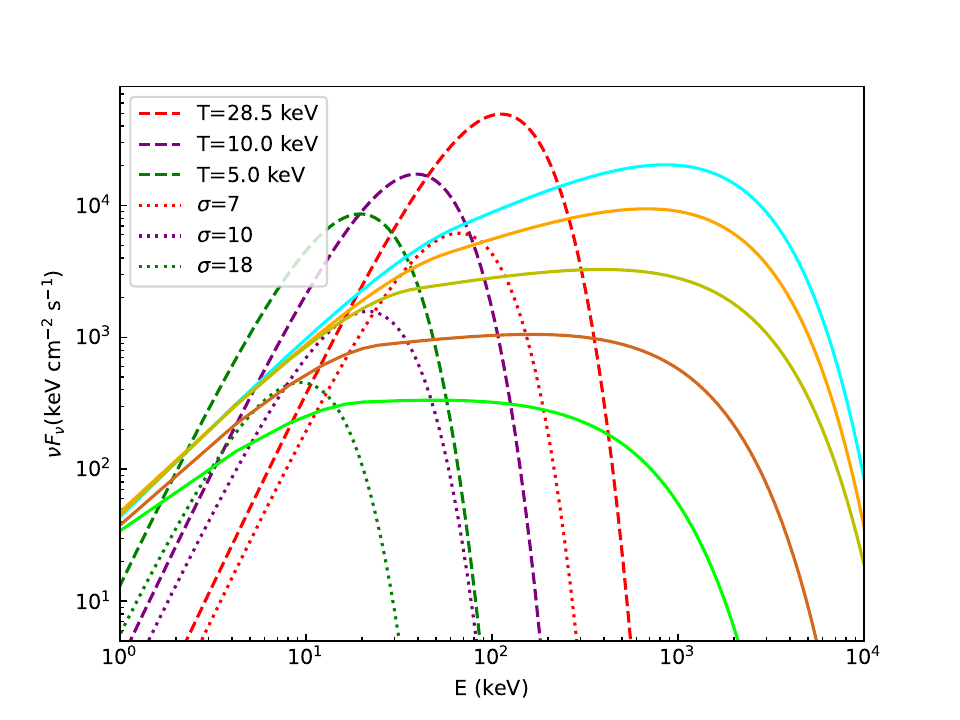}
    \includegraphics[width=3.5in]{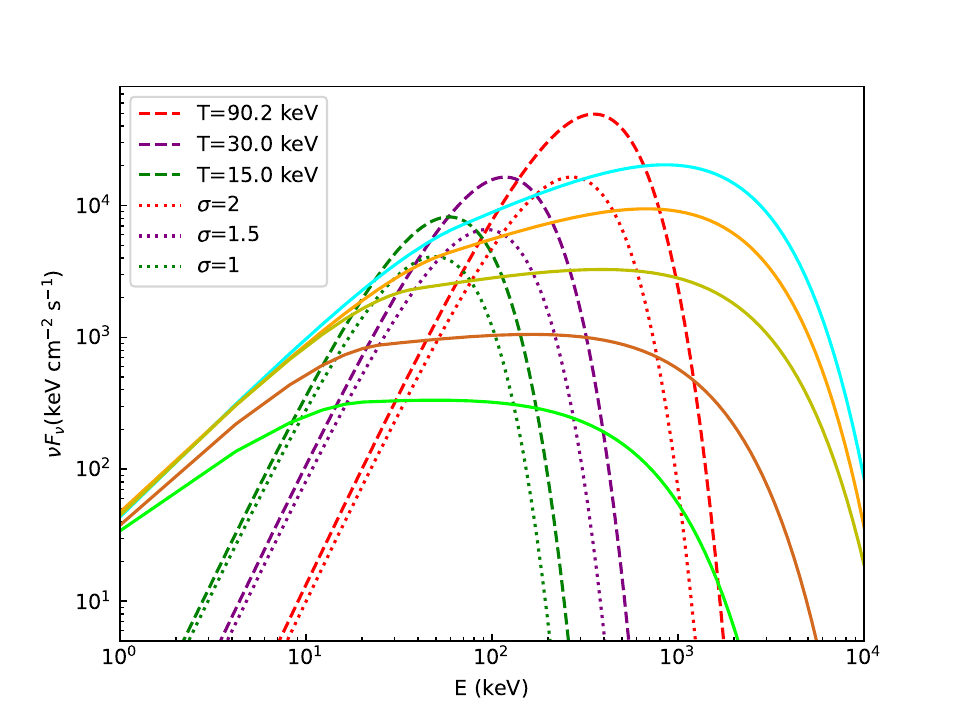}
    \caption{Left: The observed Band-cut spectra for five epochs (taken from Sun et al. (2023)) and the predicted lower limits of the photosphere spectra (dashed lines) for different temperatures with $R_0=10^{10}$ cm by adopting the epoch (a) within the framework of the baryonic fireball models. The dotted lines represent a thermal emission that makes the photosphere emission unobservable with different temperatures. Right: Similar to the left panel, but for different temperatures with $R_0=10^{9}$ cm.}
    \label{fig1}
\end{figure*}

By assuming that a pseudo blackbody spectrum is produced by the photosphere of GRB 230307A, we plot
the lower limit of the expected photosphere
spectrum for the internal shock model in the baryon-dominated outflow ($L_{\rm w}=L_{\gamma}$) in Figure \ref{fig1}. Then, we compare it with the observational data. To ensure a clear comparison, we only plot the spectra corresponding to the epoch (a) in Figure \ref{fig1}. Moreover, the pseudo thermal emission depends on the radius of central engine $R_{0}$ which is poorly understood. \cite{2023arXiv230705689S} reported that the minimum variability time scale of GRB 230307A is as short as $\delta t^{\rm ob} \sim 0.01$ s. It corresponds to radius of central engine $R_{0}=3\times 10^{8}$ cm. So, we here adopt $R_{0}=10^{9}$ cm and $R_{0}=10^{10}$ cm to do the calculations. In the left panel of Figure \ref{fig1}, the red-dashed line represents a maximum temperature with $T^{\rm ob}_{\rm ph} = T^{\rm ob}_{\rm ph,max}=28.5~\rm keV$ corresponding to $L_{\rm th}=L_{\rm w}$, $R_{\rm ph}\approx R_{\rm c}$, and $R_{0}=10^{10}$ cm. In order to compare, we also present two cases with lower temperatures for $T^{\rm ob}_{\rm ph}=10$ keV and $T^{\rm ob}_{\rm ph}=5$ keV with $\eta=60$ and $\eta=46$, respectively. Similar to the left panel of Figure \ref{fig1}, we also plot expected flux level of the photosphere emission which can be suppressed by the lower limits of magnetization parameter for different parameters within the framework of the baryonic fireball models in the right panel of Figure \ref{fig1}, but corresponding to $R_{0}=10^{9}$ cm and $T^{\rm ob}_{\rm ph} = T^{\rm ob}_{\rm ph,max}=90~\rm keV$, as well as $T^{\rm ob}_{\rm ph}=$30 keV and 15 keV with $\eta=$105 and 81, respectively.

In the left panel of Figure \ref{fig1}, we find that the pseudo thermal emission is significantly higher than that of observed non-thermal emission by adopting $R_{0}=10^{10}$ cm, and it is strongly in contradiction to the observed non-thermal Band-Cut component in the prompt emission of GRB 230307A. The contradiction strongly suggests that the baryonic model at least does not work for GRB 230307A and the initial wind luminosity is not stored in the fireball form. One possible way may be to solve the above contradiction, namely, invoking a Poynting-flux-dominated outflow \citep{2009ApJ...700L..65Z}. If this is the case, the thermal emission can be suppressed (or much dimmer), and the missing luminosity as the Poynting-flux luminosity is not observable before strong magnetic dissipation at a much larger radius \citep{2002ApJ...581.1236Z, 2002A&A...388..189D}.  In order to suppress the bright thermal emission, one can infer a lower limit on the magnetization parameter ($\sigma=L_{\rm p}/L_{\rm b}$) which is defined as the ratio between the Poynting flux ($L_{\rm p}$) and the baryonic flux ($L_{\rm b}$). The wind luminosity can be rewritten as $L_{\rm w}=L_{\rm p}+L_{\rm b}=(1+\sigma)L_{\rm b}$ \citep{2002ApJ...581.1236Z}. In the derived equations of the photosphere above (Eqs. (6), (7), and (8)), the $L_{\rm w}$ can be replaced with $L_{\rm w}/(1+\sigma)$ by assuming no dissipation of the Poynting flux below $R_{\rm ph}$. The precise value of $\sigma$ is difficult to obtain from observational data, but one can infer the minimum value of $\sigma$ which can be used to suppress (or hide) the expected thermal component from photosphere emission. 

In the left panel of Figure \ref{fig1}, the dotted lines represent a thermal emission that makes the photosphere emission unobservable with different temperatures. It is found that the required values of $\sigma$ for different temperatures at least are as high as 7, or even higher. At such high-$\sigma$, the internal shock can not power gamma-ray emission which is inconsistent with the observational data. It suggests that the internal shock model at least is not a viable mechanism to interpret GRB 230307A. So, We can confidently conclude that the ejecta of GRB 230307A is initially not in the fireball form, but is likely in a Poynting flux entrained with the baryonic matter. However, if we adopt $R_{0}=10^{9}$ cm, it is found that the pseudo thermal emission is higher a little bit than that of observed non-thermal emission, and it requires the $\sigma \sim 2$ which is not strongly enough to support the Poynting-flux-dominated outflow.

\section{Radiative efficiency of prompt emission}
For different jet compositions of GRB, they may correspond to  different energy dissipation (shocks vs. magnetic reconnection), radiation mechanisms (quasi-thermal vs. synchrotron), and dissipated radius \citep{1994ApJ...430L..93R, 2009Sci...323.1688A, 2010ApJ...709L.172R, 2009ApJ...700L..65Z, 2011ApJ...726...90Z}. The lack of detection of thermal emission of GRB 230307A implies that the photosphere model is disfavored by the energy dissipation. Theoretically, the internal shock model predicts a lower radiative efficiency (e.g., typically less than 10\%; \citealt{1999ApJ...523L.113K, 1999ApJ...522L.105P}), while the ICMART model can give a relatively high efficiency (e.g., 35\%$\sim$50\%) by invoking a runaway generation of mini-jets \citep{2007ApJ...655..989Z,2011ApJ...726...90Z,2015ApJ...805..163D}.

The GRB efficiency is defined as \citep{2004ApJ...613..477L}
\begin{equation}
    \eta_{\gamma} = \frac{E_{\gamma,\rm iso}}{E_{\rm K,iso} + E_{\gamma,\rm iso}}.
\end{equation}
where $E_{\gamma,\rm iso}$ is the isotropic energy of prompt $\gamma$-ray emission, and $E_{\rm K,iso}$ is isotropic kinetic energy. The $E_{\gamma,\rm iso}$ is usually
derived from the observed fluence ($S_{\gamma}$) in the detector’s energy band. Due to the limited detector’s energy band, we extrapolate to the rest-frame 1$–10^{4}$ keV by using spectral parameters, called $k$-correction \citep{2001AJ....121.2879B}.
\begin{equation}
    E_{\gamma,\rm iso} = 4\pi k d_{z}^2 S_{\gamma}(1 + z)^{-1}
\end{equation}
where $d_{z}$ is luminosity distance. Here, we adopt the spectral fitting results from GECAM \citep{2023arXiv230705689S} to do the k-correction due to the pulse pileup of saturation Fermi/GBM \citep{2023GCN.33407....1D}. One has $E_{\gamma,\rm iso}\sim 4\times10^{52} \rm~erg$. The isotropic kinetic energy $E_{\rm K,iso}$ term, on the other hand, there are two methods to estimate the value of $E_{\rm K,iso}$. One is used to estimate it from the afterglow data through modeling which depends on the uncertain shock microphysics parameters, such as $\epsilon_e$ and $\epsilon_B$ \citep{2001ApJ...547..922F,2007ApJ...655..989Z}. The other one is to directly calculate the $E_{\rm K,iso}$ by invoking the dominant thermal spectral component and bulk Lorentz factor \citep{2021ApJ...909L...3Z,2023ApJ...944L..57L}. In our calculations, due to the lack of thermal emission of GRB 230307A, we adopt the first method to calculate $E_{\rm K,iso}$ by using afterglow data.

Although the lack of detection of X-ray emission by Swift/XRT, fortunately, the Lobster Eye Imager for Astronomy (LEIA) caught the soft X-ray emission (0.5–4 keV) of this burst exactly at its trigger time \citep{2023arXiv230705689S}. More interestingly, the soft X-ray light curve exhibits a plateau emission followed by a normal decay phase, namely, a smoothly broken power-law fit with decay slopes $\alpha_1\sim 0.4$, $\alpha_2\sim 2.33$, and break time $t_b\sim 80$ s \citep{2023arXiv230705689S}. In order to calculate the $E_{\rm K,iso}$ based on the soft X-ray emission of GRB 230307A, one needs to judge the spectral regime and environment, e.g., interstellar medium (ISM) or wind. Following the method in \cite{2006ApJ...642..354Z} and \cite{2014ApJ...785...74L}, two independent criteria should be satisfied, namely $\alpha_1-\alpha_2$, and the “closure relation” $\alpha_2-\beta_X$. Here $\beta_X$ is the spectral index of the normal decay segment which is X-ray photon index minus 1. 

\begin{figure*}
    \centering
    \includegraphics[width=3.5in]{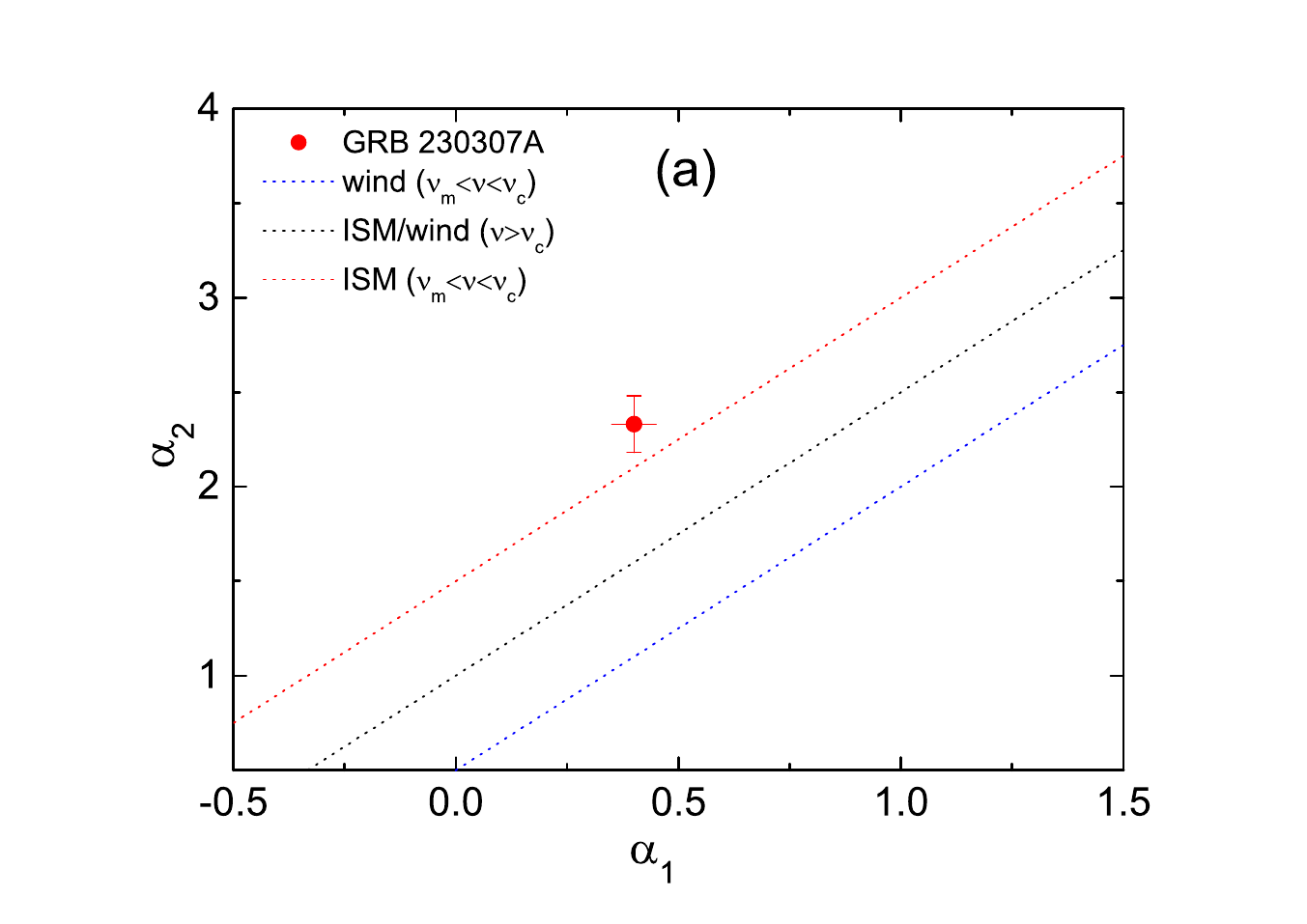}
    \includegraphics[width=2in]{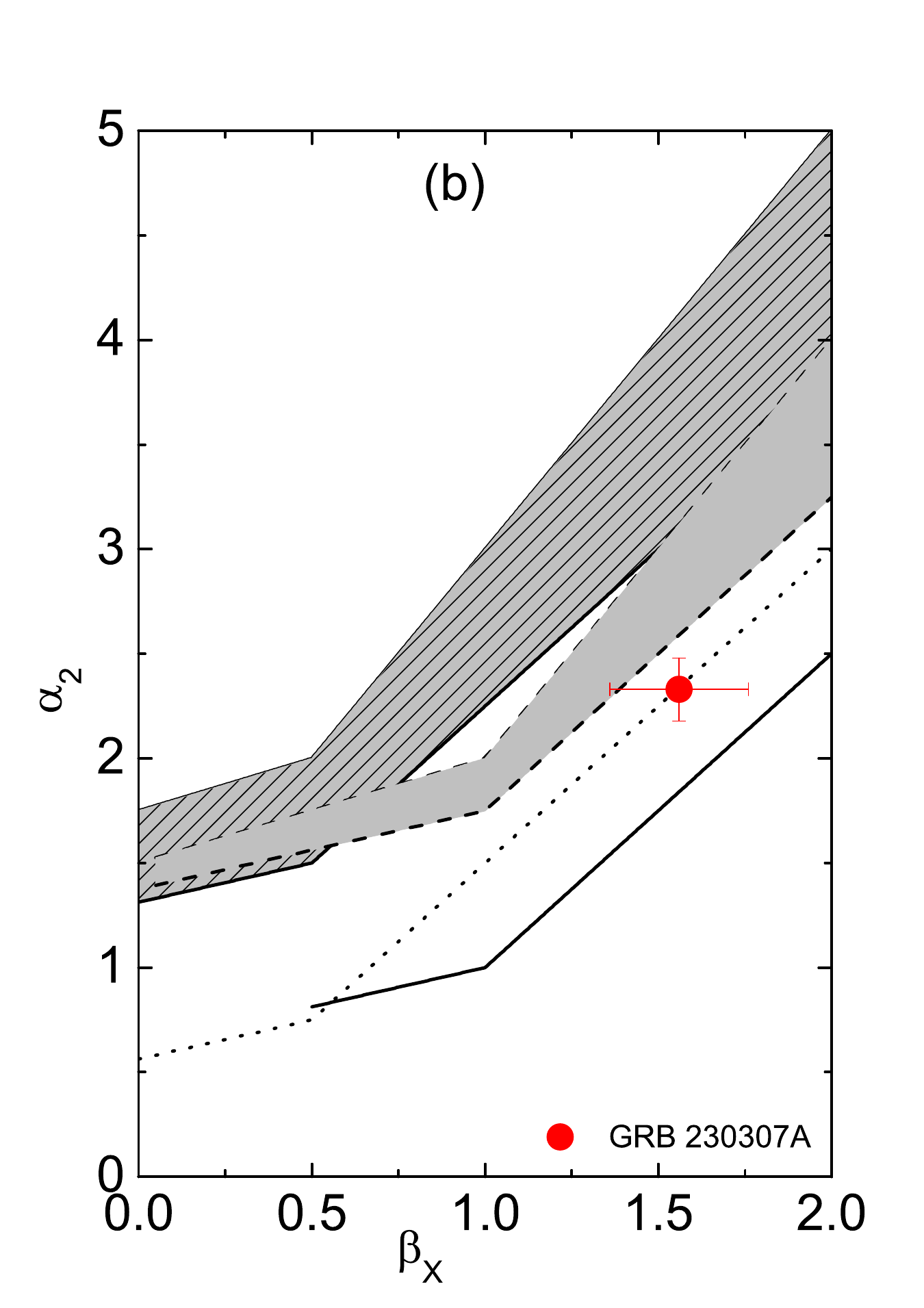}
    \caption{(a): Relationships between temporal decay indices $\alpha_1$ and $\alpha_2$ in different spectral regions and environments. The three solid lines indicate the closure relations of three specific external shock models invoking a constant energy injection. (b): Temporal decay index $\alpha_2$ against spectral index $\beta_X$ along with the closure relations of the external shock model in the case of ISM. The solid line (pre-jet break) and the shaded region (post jet break) are for the spectral regime I ($\nu>\rm max (\nu_c, \nu_m)$, while the dashed line (pre-jet break) and hatched region (post jet break) are for the spectral regime II ($\nu_m<\nu<\nu_c$).}
    \label{fig2}
\end{figure*}

\cite{2001ApJ...552L..35Z} studied energy injection from a central engine with a general luminosity law as $L(t)=L_{0}(t/t_0)^{-q}$, the pre-break slope $\alpha_1$ should correspond to a constant energy injection from a central engine (e.g., $L(t)=L_{0}(t/t_0)^{0}$), while the post-break central engine luminosity should be $L(t)=L_{0}(t/t_0)^{-2}$. For the external shock model scenario, any q steeper than -2 has no effect, and it should be treated as a constant energy in the jet as $L(t)=L_{0}(t/t_0)^{-1}$ \citep{2006ApJ...642..354Z,2013NewAR..57..141G}. So that, the relationship can be written as
\begin{equation}
\begin{split}
\alpha_{1}=\left\{
\begin{aligned}
&\frac{2\alpha_{2}-3}{3}, & \nu_{m}<\nu<\nu_{c}\rm~ (ISM)\\
&\frac{2\alpha_{2}-1}{3}, & \nu_{m}<\nu<\nu_{c}\rm~ (Wind)\\
&\frac{2\alpha_{2}-2}{3}, & \nu>\nu_{c}\rm~(ISM~or~Wind).
\end{aligned}
\right.
\end{split}
\end{equation}
Moreover, the temporal and spectral properties of the afterglow after the break (the normal decay phase) should satisfy the 'closure relation' of the external shock model \citep{2004IJMPA..19.2385Z,2007ApJ...670..565L,2013NewAR..57..141G}
\begin{equation}
\begin{split}
\alpha_{2}=\left\{
\begin{aligned}
&\frac{3\beta_X}{2}=\frac{3(p-1)}{4}, & \nu_{m}<\nu<\nu_{c}\rm~ (ISM)\\
&\frac{3\beta_X+1}{2}=\frac{3p-1}{4}, & \nu_{m}<\nu<\nu_{c}\rm~ (Wind)\\
&\frac{3\beta_X-1}{2}=\frac{3p-2}{4}, & \nu>\nu_{c}\rm~ (ISM~or~Wind)
\end{aligned}
\right.
\end{split}
\end{equation}
Here, $p$ is the electron’s spectral distribution index. Based on the spectral fitting results from \cite{2023arXiv230705689S} that is $\beta_X\sim 1.56$, one can easily judge that it should be located in the spectral regime $\nu_{m}<\nu<\nu_{c}$ and ISM (see Figure \ref{fig2}).

Based on the method in \cite{2007ApJ...655..989Z}, the $E_{\rm K,iso}$ can be calculated as below in ISM with $\nu_{m}<\nu<\nu_{c}$, 
\begin{equation}
\begin{aligned}
    E_{\rm K,iso,52}= &\Big[  \frac{\nu F_{\nu}(\nu=10^{18}\rm Hz)}{5.2\times 10^{-14}{\rm erg~s^{-1}~cm^{-2}}}   \Big]^{4/(p+2)}\\ 
    &\times d_{z,28}^{8/(p+2)}(1+z)^{-1}t_d^{(3p-2/(p+2)}\\
    &\times (1 + Y)^{4/(p+2)}f_p^{-4/(p+2)}\epsilon_{B,-2}^{(2-p)/(p+2)}\\
    &\times\epsilon_{e,-1}^{4(1-p)/(p+2)}\nu_{18}^{2(p-2)/(p+2)}
\end{aligned}
\end{equation}
where $\nu F_{\nu}(\nu=10^{18}\rm Hz)$ is the energy flux at $10^{18}\rm Hz$, $t_d$ is the time in the observer frame in days. Initially, in our calculations, the microphysics parameters of the shock $\epsilon_e$, $\epsilon_B$, and Compton parameter $Y$, we adopt the typical values derived from observations $\epsilon_B=0.01$, $\epsilon_e=0.1$, and $Y=1$, respectively \citep{2002ApJ...571..779P, 2003ApJ...597..459Y}. However, the kinetic energy of the afterglow depends on both $\epsilon_e$ and $\epsilon_B$, more sensitively with $\epsilon_e$. Then, we adopt different values of $\epsilon_e=0.1$ and 0.01, and $\epsilon_B=0.01, 0.001$, and 0.0001 to calculate the kinetic energy and efficiency (see Table 1). $f_p$ is a function of the power law distribution electron spectral index $p$ \citep{2007ApJ...655..989Z} 
\begin{equation}
    f_p \sim 6.73\Big(\frac{p-2}{p-1}  \Big)^{p - 1} (3.3\times 10^{-6})^{(p - 2.3)/2}
\end{equation}

After deriving both $E_{\gamma,\rm iso}$ and $E_{k,\rm iso}$, one can calculate the radiative efficiency for different $\epsilon_e$ and $\epsilon_B$, and the results are shown in Table 1. More interestingly, it is found that the radiative efficiency $\eta_{\gamma}$ is as high as $50\%$ of GRB 230307A for $\epsilon_e=0.1$ and $\epsilon_B=0.01$. It is significantly higher than the typical efficiency of the internal shock scenario, but is consistent with that of expected by ICMART with a Poynting-flux-dominated jet. This is also consistent with that of results from \cite{2023arXiv231007205Y} based on the mini-structure of the prompt emission lightcurve to claim the ICMART model a Poynting-flux-dominated outflow.

\section{Conclusion and Discussion} \label{sec:end} 
In this paper, we investigate the jet composition of GRB 230307A by invoking two independent methods. First, the lack of detection of thermal emission expected from the baryonic model provides strong evidence to constrain the composition of the fireball: the outflow of GRB 230307A should be Poynting-flux-dominated with a magnetization parameter ($\sigma>7$) at the photosphere in order to suppress thermal component and account for the observed spectra for $R_0=10^{10}$ cm (see Figure \ref{fig1}). It suggests that a significant fraction of the Poynting flux energy is not likely to be directly converted to the kinetic energy of the outflow below the photosphere. Second, the radiative efficiency $\eta_{\gamma} \sim 50\%$ of GRB 230307A for $\epsilon_e=0.1$ and $\epsilon_B=0.01$, is significantly higher than the typical efficiency of the internal shock scenario, but is consistent with that of expected by ICMART with a Poynting-flux-dominated jet.

In our calculations, first, we only account for epoch (a) with the brightest one to present the pseudo thermal emission and calculate the minimum values of $\sigma$ for different temperatures. Moreover, we also attempt to apply for epochs (b), (c), (d), and (e), and find that the peak of the thermal component still exceeds the observed data. It still requires that a minimum of $\sigma$ is as high as 5 (or 4, 4, and 3) for epoch (b) (or (c), (d), and (e)) with different temperatures at $R_0=10^{10}$ cm. It means that at least 83$\%$ (or 80$\%$, 80$\%$, and 75$\%$) of the energy is stored in the form of Poynting flux for epoch (b) (or (c), (d), and (e)) even considering the evolution of jet composition during such short time-scale. Second, the calculated efficiency is dependent on the microphysical parameters of the shock, namely $\epsilon_e$ and $\epsilon_B$ which are poorly constrained from the observational data. The high efficiency $\eta_{\gamma}\sim50\%$ for typical values of $\epsilon_e=0.1$ and $\epsilon_B=0.01$, at least tell us the possible of existing high efficiency that is consistent with Poynting-flux-dominated outflow.

In any case, the lack of detection of thermal emission required a high magnetization parameter, together with a high radiation efficiency, strongly suggest that a significant fraction of the outflow energy is likely in a Poynting flux entrained with the baryonic matter rather than a pure baryonic flux. It is consistent with that of results from \cite{2023arXiv231007205Y} who claimed a Poynting-flux-dominated outflow with ICMART model based on the mini-structure of prompt emission lightcurve of this burst. 

Based on the previous studies and some observational clues of GRB 230307A, e.g., lack of supernova-associated, but associated with a possible kilonova emission, the heavy element nucleosynthesis, and the characteristic of soft X-ray emission \citep{2023ApJ...954L..29D, 2023arXiv230702098L, 2023arXiv230800638Y, 2023arXiv230705689S,2023arXiv231007205Y}, point towards a compact star merger origin, and a unified picture can be accounted for as follows. The central engine of GRB 230307A is a millisecond magnetar which is the post-merger product of NS-NS mergers (supported by a lanthanide-rich kilonova). It can launch a relativistic jet via energy extraction mechanisms, such as spin-down, magnetic bubble eruption due to differential rotation, or accretion \citep{2018pgrb.book.....Z}. The observed bright gamma-ray emission can be powered by dissipation of the Poynting flux energy via synchrotron radiation. The extra soft X-ray plateau component detected by LEIA \citep{2023arXiv230705689S} can be powered by magnetic dipole spin-down of magnetar. Finally, the shocks interact with the interstellar medium to produce the observed multi-wavelength afterglow emission. However, we can not completely rule out the NS-BH merger (or NS-NS merger) with the BH central engine.

\section*{Acknowledgements}
This work is supported by the Guangxi Science Foundation the National (grant Nos. 2023GXNSFDA026007, and 2017GXNSFFA198008), the Natural Science Foundation of China (grant Nos. 11922301 and 12133003), and the Program of Bagui Scholars Program (LHJ).

\section*{Data Availability}
This is the theoretical work, and there are no new data associated with this article.


%%%%%%%%%%%%%%%%%%%%%%%%%%%%%%%%%%%%%%%%%%%%%%%%%%
%%%%%%%%%%%%%%%%%%%%%%%%%%%%%%%%%%%%%%%%%%%%%%%%%\
\begin{table*}
    \centering
    \caption{Radioactive efficiency for different values of $\epsilon_B$ and $\epsilon_e$.}
    \label{table1}
\begin{tabular}{ccc}
    \hline
    \hline
    $\epsilon_e$ & $\epsilon_B$  &  $\eta_{\gamma}$ \\
    \hline
      0.1  &   0.01   &   50.7\% \\
      0.1  &  0.001   &  31.66\% \\
      0.1  &  0.0001  &  17.26\%  \\
      \hline
      0.01 &  0.01   &  0.93\%   \\
      0.01 &  0.001  &  0.42\% \\
      0.01  &   0.0001  &  0.19\% \\
      \hline
\end{tabular}
\end{table*}
%%%%%%%%%%%%%%%%%%%%%%%%%%%%%%%%%%%%%%%%%%%%%%%%%%
%%%%%%%%%%%%%%%%%%%%%%%%%%%%%%%%%%%%%%%%%%%%%%%%%%
%\clearpage
%**************************************************

%%%%%%%%%%%%%%%%% APPENDICES %%%%%%%%%%%%%%%%%%%%%
%%%%%%%%%%%%%%%%%%%%%%%%%%%%%%%%%%%%%%%%%%%%%%%%%%

% Don't change these lines
\bsp	% typesetting comment
\label{lastpage}
\end{document}